\documentclass[prl,twocolumn]{revtex4}
\usepackage{amssymb}

\input{tcilatex}

\begin{document}

\title{Is the spectrum of highly excited mesons purely coulombian?}
\author{El Houssine Mezoir}
\author{P. Gonz\'alez}
\affiliation{Departamento de F\'{\i}sica Te\'{o}rica, Universidad de Valencia (UV) and
IFIC (UV-CSIC), Valencia, Spain.}
\date{\today}

\begin{abstract}
We show that a static central potential may provide a precise description of
highly excited light unflavoured mesons. Due to string breaking this
potential becomes of chromoelectric type at sufficiently large
quark-antiquark distances giving rise to a coulombian spectrum. The same
conclusion can be inferred for any other meson sector through a
straightforward extension of our analysis.
\end{abstract}

\maketitle

In the last years the interest in the highly excited light-quark meson
spectrum has been renewed as a consequence of the observation of more than
thirty new meson resonances with masses between $1.9$ and $2.4$ GeV in the
exhaustive analyses of Crystal Barrel and PS172 data \cite{Bug04}. Many of
the reported new resonances are listed by the Particle Data Group (PDG) \cite%
{PDG06} in the section \textquotedblleft Other Light Unflavoured
Mesons\textquotedblright\ awaiting for confirmation from a separate
experiment. Regarding the non-new resonances also found in \cite{Bug04} they
are in perfect correspondence (quite similar masses) with the ones listed in
the \textquotedblleft Light Unflavoured Mesons\textquotedblright\ section of
the PDG. From the theoretical point of view the extensive spectrum up to $%
2.4 $ GeV has made clear an approximate hydrogen like classification of
meson states \cite{Afo07}. In this article we show through a dynamical study
that the physical origin of the hydrogen like degeneracy may have to do with
string breaking once the quark-antiquark average distance in a meson reaches
a sufficiently high value. We shall rely on a Constituent Quark Model (CQM)
calculation by solving the Schr\"{o}dinger equation for a static potential.
Although the use of a static potential for light-quark systems is in general
debatable we shall justify its applicability for highly excited states. Then
in the spirit of CQM calculations we shall assume that relativistic
corrections to the Schr\"{o}dinger equation may be to a good extent taken
into account through the effective character of the parameters of the
potential.

From lattice QCD the static potential is the ground state energy of a bound
state composed of two static colour sources and gluons \cite{Bal01}. In the
so called quenched approximation (only valence quark ($q)$ and antiquark ($%
\overline{q}$)) the static $q\overline{q}$ potential, given by the
expectation value of the Wilson loop operator, resembles a funnel potential
containing a confining term depending linearly on $r,$ the $q$-$\overline{q}$
distance, plus a coulombian term proportional to $(1/r).$ When including sea
quarks, i.e. in the so called unquenched approximation, the static sources $%
q $ and $\overline{q}$ are screened by light quark-antiquark pairs that pop
out of the vacuum. Then transitions between the static sources state
(string) and mesons coming from the recombination of the static sources and
the members of the pairs may take place. Two physical effects occur. On the
one hand the confining term becomes a constant (string breaking) at a
certain saturation distance, $r_{s}.$ Although this behaviour has not been
detected with the Wilson loop technique, the finite temperature potential
extracted from Poliakov line correlators at temperatures close to the
deconfinement phase transition, exhibits a flattening, once sea quarks are
included into the action \cite{DeT98}. Such a flattening occurs at distances 
$r_{s}\simeq 1.15$ fm$.$The diagonalization of a two by two correlation
matrix between the string state and a two meson state (each meson formed by
one static source and one member of a pair) should confirm it. On the other
hand the coulomb strength remains higher than in the quenched case. From
these indications we shall assume a static potential of the form%
\begin{eqnarray}
V(r) &=&\sigma r-\frac{k}{r}+C\text{ \ \ \ \ \ \ \ \ \ \ \ \ \ if \ \ }r\leq
r_{s}  \label{sp} \\
V(r) &=&\sigma r_{s}-\frac{k}{r}+C\text{ \ \ \ \ \ \ \ \ \ \ \ \ if \ \ }%
r\geq r_{s}  \nonumber
\end{eqnarray}%
where $\sigma $ stands for the string tension, $k$ for the coulomb strength
and $C$ for a constant to fix the origin of the potential.

We shall choose this form for the effective $q\overline{q}$ potential in our
CQM calculation. Let us recall that in quark models the effective $q%
\overline{q}$ potential is found by equating the scattering amplitude of
free quark and antiquark with the potential between bound quark and
antiquark inside a meson. Thinking of a single exchange diagram the static
limit corresponds to no energy transfer,\textit{\ }i.e. to $q^{0}\equiv
\left( E_{q}\right) _{initial}-\left( E_{q}\right) _{final}=\left( E_{%
\overline{q}}\right) _{final}-\left( E_{\overline{q}}\right) _{initial}=0.$
So the static approximation means $q^{0}\simeq 0.$ Then by substituting $%
E_{q}=m_{q}\sqrt{1+(\overrightarrow{p}_{q}^{2}/m_{q}^{2})}$ we can easily
establish as a criterium for the validity of such approximation the
requirement%
\begin{equation}
\frac{\left[ \left( \overrightarrow{p}_{q}^{2}\right) _{initial}-\left( 
\overrightarrow{p}_{q}^{2}\right) _{final}\right] }{m_{q}^{2}}<<1
\label{crit2}
\end{equation}%
By replacing $\left( \overrightarrow{p}_{q}\right) _{initial}=%
\overrightarrow{q}+\left( \overrightarrow{p}_{q}\right) _{final}$ the
criterium~(\ref{crit2}) can be written as 
\begin{equation}
\frac{\overrightarrow{q}^{2}+2\overrightarrow{q}.\left( \overrightarrow{p}%
_{q}\right) _{final}}{m_{q}^{2}}<<1  \label{crit3}
\end{equation}%
For non-relativistic as well as for relativistic systems~(\ref{crit3}) is
satisfied when 
\begin{equation}
\frac{\left\vert \overrightarrow{q}\right\vert }{m_{q}}<<1\text{ \ \ \ \ \ \
and \ \ \ \ \ \ \ }\frac{\left\vert \overrightarrow{p}_{q}\right\vert }{m_{q}%
}\ngtr \ngtr 1  \label{cond}
\end{equation}%
Since the main contributions to the interaction come from distances $%
r\approx \hbar c/\left\vert \overrightarrow{q}\right\vert $ we expect the
mesons satisfying~(\ref{cond}) to have root mean square radius (rms-radius) 
\begin{equation}
<r^{2}>^{1/2}\simeq \frac{1}{\left\vert \overrightarrow{q}\right\vert }>>%
\frac{1}{m_{q}}  \label{radi}
\end{equation}%
Notice that for bottomonium ($b\overline{b})$ and charmonium $(c\overline{c})
$ the conditions~(\ref{radi}) and $\left\vert \overrightarrow{p}%
_{q}\right\vert /m_{q}\ngtr \ngtr 1$ are well satisfied according to
reference~\cite{Eic78} where a good description of such mesons is attained
by means of a quenched potential. The validity of the quenched approximation
in this case can be related to the fact that for $b\overline{b}$ and $c%
\overline{c}$ the saturation distance $r_{s}$ may be significantly larger
than $1.15$ fm. The screening effect on $b\overline{b}$ and $c\overline{c}$
has been studied in the literature \cite{Din95}. From these studies a value
for $(r_{s})_{b\overline{b},c\overline{c}}$ up to $1.8$ fm may be
conjectured. This value may be reflecting the fact that the $b\overline{b}$
and $c\overline{c}$ decays into stable hadrons involve more than two mesons.
Then it can be explained why the predicted $b\overline{b}$ and $c\overline{c}
$ spectra involving states with $<r^{2}>^{1/2}\lesssim (r_{s})_{b\overline{b}%
,c\overline{c}}$ hardly change from the unquenched to the quenched
approximation. The experimental extension of the spectra to states with
larger rms-radii is essential to confirm or refute this conjecture.

Here we centre on light-quark mesons. We shall restrict for simplicity to
isospin $I=1$ mesons (they contain only $u$ ($\overline{u})$ and/or $d$ ($%
\overline{d}))$: $\pi _{J},$ \textbf{b}$_{J}$, $\rho _{J}$ and \textbf{a}$%
_{J}$ ($J:$ total angular momentum). Thus we avoid all possible
complications coming from $q\overline{q}$ annihilation and from $s\overline{s%
}$ components. The mass of the constituent quarks $m_{u(\overline{u}),\text{ 
}d(\overline{d})},$ named henceforth $m_{u}$, is a parameter of our model.
We shall fix its value from the average dynamical quark mass generated by
Spontaneous Chiral Symmetry Breaking (SCSB), $m_{u}(\left\vert 
\overrightarrow{q}\right\vert )$, in the energy region under consideration.
From instanton model calculations~\cite{Dia02} confirmed by lattice QCD~\cite%
{Bow04} we know the explicit $m_{u}(\left\vert \overrightarrow{q}\right\vert
)$ dependence so that it has its maximum value at $\left\vert 
\overrightarrow{q}\right\vert =0$ ($m_{u}(0)\simeq 0.350$ GeV) and decreases
when increasing $\left\vert \overrightarrow{q}\right\vert $. For $\left\vert 
\overrightarrow{q}\right\vert =0.1$ GeV for instance one has $m_{u}(0.1$ GeV$%
)=0.332$ GeV. Therefore~(\ref{cond}) tells us that the static approximation
might only be applied for $\left\vert \overrightarrow{q}\right\vert <0.1$ GeV%
$.$ Then we shall use $m_{u}=340$ MeV as an average mass in this interval
for which we expect from~(\ref{radi}) to have mesons with $%
<r^{2}>^{1/2}>>0.6 $ fm. It should be added that SCSB has another important
effect: the appearance of Goldstone bosons. This effect is not explicitly
reflected in~(\ref{sp}). We shall comment on this later on.

To check the applicability of the static approximation we proceed to
calculate the spectrum of $I=1$ mesons with an effective potential of the
form~(\ref{sp}). For this purpose we have to fix the parameters of the
potential. For $r_{s}$ we take $(r_{s})_{u}=1.15$ fm. For the string tension 
$\sigma _{u}$ we shall use the value $\sigma _{u}=932.7$ MeV/fm obtained
from the Regge trajectory for $\rho _{J}$ and \textbf{a}$_{J}$ (see for
instance \cite{Bal01}) in accord with lattice evaluations. Regarding $k_{u}$
and $C_{u}$ we shall fix them by fitting the average masses of the
experimental states with higher orbital angular momentum $L,$ since we
expect these states to have large rms-radii due to the centrifugal barrier.
Let us realize that the calculated masses coming out from the Schr\"{o}%
dinger equation will only depend on $L$ ($L=0,1,2,3...)$ and on the radial
quantum number $n_{r}$ ($n_{r}=1,2,3...).$ We shall denote them as $%
M_{L,n_{r}}.$ So we should compare them to the average masses of
experimental $(L,n_{r})$ multiplets, (($M)_{L,n_{r}})_{Exp}$. The existence
of such multiplets has been suggested elsewhere \cite{Afo07}. Actually the
assumption of a long distance interaction depending only on $r$ drives
naturally to a $SU(4)_{Spin-Isospin}\times O(3)$ group of symmetry so that
the $I=1$ light unflavoured mesons belong to $15$-plets (note that the
product of $SU(4)$ quark and antiquark representations is $4\times \overline{%
4}=15+1;$ the singlet representation $1$ contains only $I=0$ mesons that we
do not consider$).$ Indeed we should better talk about super $15$-plets
since for each member of the multiplet there are so many experimental states
as possible different $J$ values ($J$ degeneracy). As we consider only $I=1$
mesons we define their average mass in the corresponding $15$-plet $%
(L,n_{r}) $ as%
\begin{equation}
(M_{L,n_{r}})_{Exp}=\left( \sum\limits_{J}(2J+1)\right)
^{-1}\sum\limits_{X,J}(2J+1)X_{J}  \label{aver}
\end{equation}%
being $X_{J}$ the experimental masses assigned to the multiplet ($%
X_{J}\equiv M_{\pi _{J}},$ $M_{\rho _{J}}$ or $M_{\text{\textbf{b}}_{J}},$ $%
M_{\text{\textbf{a}}_{J}}$).

The maximum value of $L$ for which we have some candidate from the PDG
catalog (the section \textquotedblleft Light Unflavoured
Mesons\textquotedblright ) or from reference \cite{Bug04}, named henceforth
CBC (for Cristal Barrel Collaboration), is $L=5.$ In fact there is only one
candidate, the PDG resonance \textbf{a}$_{6}(2450\pm 130).$ Although the
error bar is big the existence of a non considered $I=0$ PDG resonance 
\textbf{f}$_{6}(2465\pm 50)$ that can be assigned to the same $15$-plet ($%
L=5,n_{r}=1)$ makes us confident about the average PDG mass. For ($L=4,$ $%
n_{r}=1)$ we also have only one PDG resonance, $\rho _{5}(2330\pm 35)$ but a
complete set of CBC candidates ($\pi _{4}(2250\pm 15),$ $\rho _{3}(2260\pm
20),$ $\rho _{4}(2230\pm 25),$ $\rho _{5}(2300\pm 45))$ (an explanation for
the PDG - CBC difference in mass for $\rho _{5}$ is given in reference \cite%
{Bug04}). Let us also notice the existence of $I=0$ CBC resonances, $\omega
_{3},$ $\omega _{4}$ and $\omega _{5}$ at about the same mass.

By choosing $k_{u}=2480$ MeV.fm and $C_{u}=1070$ MeV we reproduce correctly
their average masses (for $(L=4,$ $n_{r}=1)$ a value in between the PDG and
CBC averages is chosen). It is noteworthy that the calculated rms-radii for
the fitting states ($3.7$ fm and $2.6$ fm) consistently satisfy $%
<r^{2}>^{1/2}>>0.6$ fm. Moreover the calculated values of $\left\vert 
\overrightarrow{p}_{q}\right\vert /m_{q}$ ($1.05$ and $1.3)$, although
indicating the relativistic character of quark and antiquark, are not much
greater than $1$, as required from~(\ref{cond}).

The results for the spectrum of states with $<r^{2}>^{1/2}>2.0$ fm, for
which we expect the static approximation may work (all the states have $%
\left\vert \overrightarrow{p}_{q}\right\vert /m_{q}<1.7)$, are shown in
Table~\ref{t1} (we include for completeness multiplets with rms radii below $%
2.0$ fm as ($2,2$) and $(3,1))$. Our results are compared to CBC and PDG
average masses. The states entering in the calculation of the averages are
specified. The multiplets ($1,4$) and ($1,3)$ lack an \textbf{a}$_{0}$
despite having an available candidate \textbf{a}$_{0}(2025).$ The reason for
not including this state is, appart from the ambiguity in its assignment,
the general deficient description of a$_{0}$ states provided by quark models
pointing out the need to incorporate more than two valence components. It
should be noted additionally that $\rho _{3}(2260\pm 20)$ has been assigned
to two multiplets (4,1) and (2,3). The reason for this double assignment is
the assumption that the $\rho _{3}$ resonances belonging to such multiplets
would be, as indicated by their partners in the multiplets, almost
degenerate. All the multiplets considered have at least one PDG cataloged
state or one resonance rated at least three stars in reference \cite{Bug04}.
The multiplet (0,4) has not been considered since the only trustable
assignment to it, the CBC $\pi (2070)$, has only a two-star rating (let us
point out that if we assigned the PDG $\rho (1900)$ to this multiplet our
result would be in perfect accord with data). 
\begin{table}[tbp]
\begin{tabular}{|c|c|c|c|c|}
\hline
($L$, $n_{r}$) & $<r^{2}>^{1/2}$ & $M_{L,n_{r}}$ & ($(M)_{L,n_{r}})_{CBC%
\text{ }}$ & ($(M)_{L,n_{r}})_{PDG\text{ }}$ \\ 
& fm & MeV & MeV & MeV \\ \hline
(5,1) & 3.7 & 2450$^{\dagger }$ &  & $2450\pm 130$ \\ 
&  &  &  &  \\ 
&  &  &  & a$_{6}(2450)$ \\ \hline\hline
(1,4) & 3.4 & $2255$ & $2219\pm 43$ &  \\ 
&  &  & b$_{1}(2240)$ &  \\ 
&  &  & a$_{1}(2270),$a$_{2}(2175)$ &  \\ \hline\hline
(2,3) & $3.2$ & $2254$ & $2248\pm 37$ &  \\ 
&  &  & $\pi _{2}(2245),\rho (2265)$ &  \\ 
&  &  & $\rho _{2}(2225),\rho _{3}(2260)$ & $\rho _{3}(2250)$ \\ \hline\hline
(3,2) & $2.9$ & $2258$ & $2258\pm 38$ &  \\ 
&  &  & b$_{3}(2245),$a$_{2}(2255)$ &  \\ 
&  &  & a$_{3}(2275),$a$_{4}(2255)$ &  \\ \hline\hline
(4,1) & $2.6$ & $2283^{\dagger }$ & $2262\pm 28$ & $2330\pm 35$ \\ 
&  &  & $\pi _{4}(2250),\rho _{3}(2260)$ &  \\ 
&  &  & $\rho _{4}(2230),\rho _{5}(2300)$ & $\rho _{5}(2350)$ \\ \hline\hline
(1,3) & $2.1$ & $1919$ & $1947\pm 47$ &  \\ 
&  &  & b$_{1}(1960)$ &  \\ 
&  &  & a$_{1}(1930),$a$_{2}(1950)$ &  \\ \hline\hline
(2,2) & $1.9$ & $1913$ & $1980\pm 23$ &  \\ 
&  &  & $\pi _{2}(2005),\rho (2000)$ &  \\ 
&  &  & $\rho _{2}(1940),\rho _{3}(1982)$ & $\rho _{3}(1990)$ \\ \hline\hline
(3,1) & $1.6$ & $1937$ & $2023\pm 24$ & $2010\pm 12$ \\ 
&  &  & b$_{3}(2032),$a$_{2}(2030)$ &  \\ 
&  &  & a$_{3}(2031),$a$_{4}(2005)$ & a$_{4}(2040)$ \\ \hline
\end{tabular}%
\caption{Calculated masses (column III) and rms-radii (column II) for $%
(L,n_{r})$ multiplets (column I) from the set of parameters $m_{u}=340$ MeV, 
$\protect\sigma _{u}=932.7$ MeV/fm, $k_{u}=2480$ MeV.fm and $C_{u}=1070$
MeV. Experimental CBC and PDG average masses (columns IV and V) are shown
for comparison. In both cases the candidates to be members of the multiplets
are indicated. The superindex $\dagger $ in the (5,1) and (4,1) calculated
masses indicates the average mass values chosen in the corresponding
multiplets to fix the parameters.}
\label{t1}
\end{table}
As can be checked our results seem to agree with data for $<r^{2}>^{1/2}>2.0$
fm. To be more precise we can rely on the approximate linearity and
equidistance of Regge trajectories satisfied by data up to a mass of $2.3$
GeV \cite{Afo07,Bug04,Ani00} with a standard Regge slope of about $1.1$ GeV$%
^{2}$. From our results for $(1,3)$, $(2,3)$ a Regge slope of $1.4$ GeV$%
^{2}, $ far above the standard value, would be obtained for the correponding 
$(L,3) $ trajectories. We interpret this as an indication that the static
approximation is doubtful for the $(1,3)$ state. Let us also note that a
Regge slope of $0.79$ GeV$^{2},$ quite below the standard value, would be
obtained from our chosen masses for $(4,1)$ and $(5,1)$ in the $(L,1)$
trajectories indicating that the calculated spectrum tends to a coulombian
one when increasing the energy. This is clearly indicated by the almost mass
degeneracy for states with the same values of $(L+n_{r})$ for $(L+n_{r})>6$ (%
$<r^{2}>^{1/2}>4.0$ fm). Then we can give a closed formula for the mass of
highly excited mesons

\begin{eqnarray}
\left( M_{L,n_{r}}\right) _{(L+n_{r})\geq 6} &\simeq &m_{u}+m_{\overline{u}%
}+\sigma _{u}(r_{s})_{u}  \nonumber \\
&-&\frac{\mu }{2}\frac{k_{u}^{2}}{(L+n_{r})^{2}}+C_{u}  \label{mass}
\end{eqnarray}%
where $\mu =m_{u}/2$ is the reduced mass of the system. The predicted masses
for multiplets with the same value of $(L+n_{r})$ are listed in Table~\ref%
{t2} up to $(L+n_{r})=9.$ Let us also remark the accidental degeneracy
between positive ($+$) and negative ($-$) parity states with $%
L(+)-L(-)=odd=n_{r}(-)-n_{r}(+).$%
\begin{table}[tbp]
\begin{tabular}{|c|c|}
\hline
($L+n_{r}$) & $m_{L,n_{r}}(MeV)$ \\ 
($L,n_{r}$) &  \\ \hline
6 & 2450 \\ 
(0,6), (1,5), (2,4), (3,3), (4,2) &  \\ \hline
7 &  \\ 
(0,7), (1,6), (2,5), (3,4) & 2548 \\ 
(4,3), (5,2), (6,1) &  \\ \hline
8 &  \\ 
(0,8), (1,7), (2,6), (3,5) & 2613 \\ 
(4,4), (5,3), (6,2), (7,1) &  \\ \hline
9 &  \\ 
(0,9), (1,8), (2,7), (3,6), (4,5) & 2657 \\ 
(5,4), (6,3), (7,2), (8,1) &  \\ \hline
\end{tabular}%
\caption{Predicted masses for some $(L,n_{r})$ multiplets with $L+n_{r}>6.$
Parameters as in Table~\protect\ref{t1}.}
\label{t2}
\end{table}
Moreover, as the coulombian energy (third term on the right hand side of~(%
\ref{mass})) has $0$ as an upper bound we predict that the $I=1$ light
unflavoured meson spectrum has an upper bound or limiting mass given by%
\begin{equation}
M_{Limit}\simeq m_{u}+m_{\overline{u}}+\sigma _{u}(r_{s})_{u}+C_{u}=2823%
\text{ MeV}  \label{limit}
\end{equation}%
This limit is compatible with current data since reported resonances with
higher mass listed in the PDG section \textquotedblleft Other Light
Unflavoured Mesons\textquotedblright\ may be assigned to mesons containing $s%
\overline{s}$. Although a specific analysis parallel to the one just
performed would be required for these states we can expect their limiting
mass to increase with respect to the value in~(\ref{limit}) by at least an
amount $\left[ (m_{s}-m_{u})+(m_{\overline{s}}-m_{\overline{u}})\right] \sim
300-500$ MeV. Then the resulting limit ($\lesssim $ $3300$ MeV) would be
compatible with all existing light unflavoured meson candidates. Beyond this
limit one meson states can not exist. Instead the system fragments into
several mesons.

Furthermore a mass limit for baryons containing only quarks $u$ and $d$ may
be derived through the simple prescription ($M_{B})_{Limit}\simeq
3m_{u}+(3/2)\left[ \sigma _{u}(r_{s})_{u}+C_{u}\right] =4213$ MeV. This
value is consistently above the most massive reported nucleon and delta
resonances in the PDG sections $N(\sim 3000$ Region) and $\Delta (\sim 3000$
Region).

With respect to the fitted values of the parameters some comment is in
order. As we do not expect significant Goldstone boson contributions for we
are dealing with $q$-$\overline{q}$ distances much larger than $M_{\pi
}^{-1} $ the parameters may be incorporating mainly gluon contributions
(apart from possible relativistic quark kinetic corrections). So the coulomb
strength, $k,$ can be tentatively related to an effective
quark-antiquark-gluon coupling, $\alpha _{s},$ through the colour relation $%
k=(4/3)\alpha _{s}.$ Then from the fitted value of $k=2480$ MeV.fm we get $%
\alpha _{s}(\left\vert \overrightarrow{q}^{2}\right\vert <0.01$ GeV$%
^{2})=9.4.$ It is interesting to realize that this value for $\alpha _{s}$
is precisely the same reported in reference \cite{Alk01} from a solution of
the truncated Schwinger Dyson equations. However this \textquotedblleft
coincidence\textquotedblright\ should be taken with caution since a smaller
value has been obtained in other calculations \cite{Agu01}. As a matter of
fact the only general conclusion we may extract from the work of all the
groups is the almost constancy of $\alpha _{s}$ for sufficiently low $%
\left\vert \overrightarrow{q}^{2}\right\vert .$ Undoubtedly, the validity
of~(\ref{sp}) for highly excited light-quark mesons has much to do with this
constancy of $\alpha _{s} $ for $\left\vert \overrightarrow{q}%
^{2}\right\vert <0.01$ GeV$^{2}.$

To summarize, our study of highly excited $I=1$ light unflavoured mesons
shows that their spectrum could be purely coulombian. As a consequence it
would have a limiting mass. In fact the idea that the meson spectrum might
have an upper bound is not new. A mass limit of $3.2$ GeV for light-quark
mesons involving $u\overline{u},$ $d\overline{d}$ as well as $s\overline{s}$
components was suggested some years ago from a study of Regge trajectories 
\cite{Bri00}. Our dynamical analysis supports such suggestion confirming
string breaking as the underlying physical mechanism. However the lack of
trustable experimental data beyond $2.5$ GeV leaves opened the door to other
interpretations. For instance one could think that the observed flattening
of the confining potential corresponded indeed to a severe softening of the
confining interaction in the region under study. Then the good effective
description achieved would be compatible with a very slight increase of the
interaction with the distance and consequently with an unbound meson
spectrum. Besides the non-relativistic hydrogen like symmetry we have made
dynamically evident may be only the effective face of a broader relativistic
symmetry in the energy region under consideration as discussed in \cite%
{Afo07}. In our model the higher the meson mass the lesser the $\left\vert 
\overrightarrow{p}_{q}\right\vert /m_{q}$ value and the less relativistic
the system (when approaching the limiting mass $\left\vert \overrightarrow{p}%
_{q}\right\vert /m_{q}\rightarrow 0)$ implying a non-relativistic coulombian
symmetry at the long range. Consistent relativistic calculations beyond our
effective CQM treatment could shed more light on this point.

Let us also emphasize that the analysis we have performed may be extended to
any other meson sector although the current lack of data makes it not
feasible. Then a definite answer to the general question about the
coulombian nature of the highly excited meson spectrum has to be postponed
until more complete data are available. We encourage an experimental effort
along this line. In the mean time we hope our results may be suggestive and
motivate further studies in the field.

This work has been partially funded by spanish MCyT and EU FEDER under
Contract No. FPA2007-65748 and by European III 506078.

\end{document}